\documentclass[aps,showpacs,twocolumn,10pt,pra]{revtex4-1}

\usepackage{amsmath}
\usepackage{amsthm}
\usepackage{amsfonts}
\usepackage{amssymb}
\usepackage{graphicx}
\usepackage{multirow}

\begin{document}
\title{Semi-Classical Wavefunction Perspective to High-Harmonic Generation}

\author{Fran\c{c}ois~Mauger$^{1}$, Paul~Abanador$^{1}$, Kenneth~Lopata$^{2}$, Kenneth~J.~Schafer$^{1}$ and Mette~B.~Gaarde$^{1}$}
\affiliation{$^{1}$Department of Physics and Astronomy, Louisiana State University, Baton Rouge, Louisiana 70803-4001, USA\\ 
$^{2}$Department of Chemistry, Louisiana State University, Baton Rouge, Louisiana 70803-4001, USA}

\date{\today}

\begin{abstract}
We introduce a semi-classical wavefunction (SCWF) model for strong-field physics and attosecond science. When applied to high harmonic generation (HHG), this formalism allows one to show that the natural time-domain separation of the contribution of ionization, propagation and recollisions to the HHG process leads to a frequency-domain factorization of the harmonic yield into these same contributions, for any choice of atomic or molecular potential.  We first derive the factorization from the natural expression of the dipole signal in the temporal domain by using a reference system, as in the quantitative rescattering (QRS) formalism [J.~Phys.~B.\ {\bf 43}, 122001 (2010)]. Alternatively, we show how the trajectory component of the SCWF can be used to express the factorization, which also allows one to attribute individual contributions to the spectrum to the underlying trajectories.
\end{abstract}



\maketitle

\section{Introduction} \label{sec:Introduction}

Building on the advances of laser technology, strong-field physics and attosecond science~\cite{Cork07,Beck08,Krau09,Beck12} have attracted a lot of attention as means to manipulate and probe the electronic structure at the atomic and molecular level~\cite{Klin08,Cork11,Lepi13,Vrak14}. Among the variety of possible outcomes from the laser-matter interaction, high harmonic generation (HHG)~\cite{Kapt07,Gaar08,Klin08} focuses on the highly nonlinear and non-perturbative process by which coherent harmonic photons of the driving laser are emitted, with harmonic orders ranging up to the extreme ultraviolet regime~\cite{Kapt07,Popm12}. In turn, the intrinsic coherence of the HHG process can be exploited in the development of novel high performance light sources such as attosecond pulses~\cite{Anto96,Klin08,Agos12}. Alternatively, by the fundamental properties of the HHG process, information on electronic structure and electron dynamics are encoded in the spectrum~\cite{Itat04,Smir09_2,Haes10,Mair10,Cork11,Krau15}, opening the way for high harmonic spectroscopy.

At the core of strong-field physics is the recollision picture~\cite{Scha93,Cork93} in which an electron, after being ionized, is accelerated and returned to its parent ion upon reversal of the electric field direction. Upon recollision, electromagnetic radiation can be emitted therefore corresponding to HHG. Following the decomposition of the process into three successive steps, one can intuitively expect the HHG cross section to factorize into the product of each individual step, as $(i)$ the ionization probability times $(ii)$ the propagation, through the probability of recollision, times $(iii)$ the efficiency of rescattering. Such a factorization has been expressed in the temporal domain for atoms~\cite{Ivan96} and was extended to include more complicated core dynamics of molecular systems~\cite{Smir09_2,Mair10}. Less intuitively, the quantitative rescattering (QSR) model has empirically shown that this factorization of the HHG spectrum can be expressed directly in the frequency domain, with results in very good agreement with full quantum simulations and experimental measurements~\cite{Le08,Lin10,Le13}. The theory for such a spectral factorization has been established for short range potentials~\cite{Frol09}. In this general context, we introduce a semi-classical wavefunction (SCWF) formalism which, by combining the wave/particle picture of the electron, leads to an intuitive derivation of the HHG spectrum factorizations, irrespective of the potential.

In most theoretical analyses and interpretations of HHG, two main approaches have been considered in the literature. On the one hand, the plane wave/Volkov state (or further refined Coulomb corrected models)~\cite{Lewe94,Ivan96,Pugl15} adopt a wave perspective of the recolliding electron. Such an approach allows one to define the recollision dipole element~[see, e.g., Eq.~(\ref{eq:HHG_spectrum})]. However, since the electronic wavefunction is completely delocalized in configuration space, at all times the ionized part of the wavefunction overlaps completely with the bound part and the instant of recollision, i.e., step $(iii)$ in the recollision picture, has to be imposed by hand. On the other hand, classical (or semi-classical) interpretations that make use of electronic trajectories~\cite{Sali01,Mair03,Milo06,Host10}, e.g., using the stationary phase approximation, allow for an intuitive definition of the recollision time but lose the dipole recollision counterpart that now has to be to some extent artificially imposed. In this paper we introduce a semi-classical wavefunction that  combines the wave/particle pictures with a quantum-like delocalized wavefunction supported by a trajectory in phase space. These combined perspectives allow us to overcome the aforementioned difficulties and naturally define the dipole signal associated with recollision.

The Article is organized as follows: 
\emph{Section~\ref{sec:Model}} defines the quantum framework in which HHG simulations are performed throughout the paper. This section also describes the reduced dimensional molecular model we use as an illustration.
\emph{Section~\ref{sec:SCWF}} defines the theoretical framework for the SCWF approximation. First we discuss the semi-classical trajectory component of the SCWF (section~\ref{sec:SCWF:Trajectory}). Then we focus on the bound part of the wavefunction and ionization step (section~\ref{sec:SCWF:Bound_state_and_ionization}). Finally, we put all these elements together to approximate the dipole acceleration signal (section~\ref{sec:SCWF:Dipole_signal}) from which HHG spectra are computed.
\emph{Section~\ref{sec:HHG_factorization}} uses the SCWF picture to derive a factorization of the HHG spectrum as the product of the ionization, propagation and rescattering $(i)\times(ii)\times(iii)$ terms, in the energy (frequency) domain. First we consider the factorization when the propagation part is described with a reference system (section~\ref{sec:HHG_factorization:Reference_system}). Then, we investigate the factorization when the semi-classical trajectory picture of the SCWF is used for the propagation term (section~\ref{sec:HHG_factorization:Direct_factorization}); This allows us to compare the relative importance of the three step cross-sections in the overall HHG spectrum.
\emph{Section~\ref{sec:Conclusion_and_perspectives}} concludes the paper and discusses some possible perspectives unveiled by the SCWF picture.

\section{Model} \label{sec:Model}

\begin{figure}
	\centering
		\includegraphics[width=\linewidth]{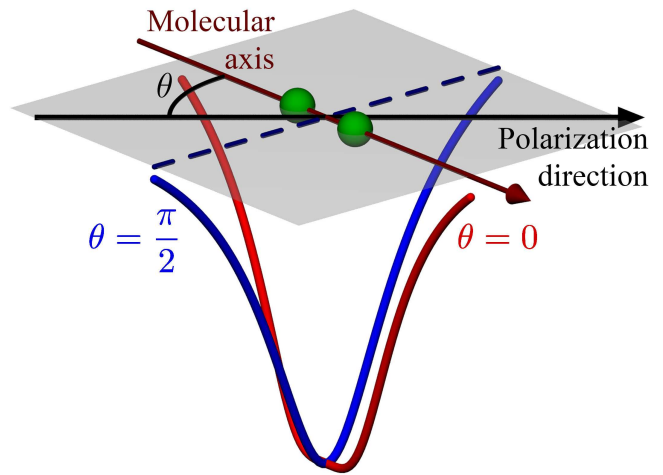}
	\caption{\label{fig:Potential_illustration} (color online)
	Illustration of the one dimensional model of potential~(\ref{eq:Potential}). The electron dynamics is restricted along the polarization direction which forms an angle $\theta$ with the molecular axis as shown in the upper part of the figure. In the lower part, we display the effective potential shapes for the two limiting angle $\theta=0$ and $\theta=\pi/2$ (used as the reference, see text) as labeled on the figure.}
\end{figure}

In this paper we consider high harmonic generation (HHG) as obtained from numerical integration of the time-dependent Schr\"{o}dinger equation (TDSE) $i\partial_{t}\left|\psi\right\rangle=\hat{\mathcal{H}}\left|\psi\right\rangle$, using bracket notations where appropriate, for an isolated single active electron (SAE) model. In the length gauge and using atomic units (unless otherwise specified) the Hamiltonian operator reads
\begin{equation} \label{eq:Hamiltonian}
	\hat{\mathcal{H}}\left(x,t\right) = 
		\hat{\mathcal{H}}_{0}\left(x\right) + \mathcal{E}\left(t\right)\hat{x} = 
		-\frac{\Delta}{2} + \mathcal{V}\left(x\right) + \mathcal{E}\left(t\right)\hat{x},
\end{equation}
where $\mathcal{V}$ is the (SAE) effective potential, $\mathcal{E}\left(t\right)$ is the laser electric field and we consider a one dimensional configuration for the sake of simplicity, as illustrated in Fig.~\ref{fig:Potential_illustration}. From the solution of the TDSE we define the associated HHG spectrum as the Fourier transform of the dipole acceleration
\begin{equation} \label{eq:HHG_spectrum}
	R_{\rm HHG}\left(\nu\right) = \mathcal{F}\left[\ddot{d}\left(t\right)\right]\left(\nu\right) \ \ \ {\rm with} \ \ \ 
	d = \left\langle\psi\left|\hat{x}\right|\psi\right\rangle,
\end{equation}
where the Fourier operator $\mathcal{F}$ is defined as 
\begin{equation} \label{eq:Fourier_transform}
	\mathcal{F}\left[f\left(x\right)\right]\left(\nu\right) = \frac{1}{\sqrt{2\pi}}\int_{\mathbb{R}}{
		dx\ f\left(x\right){\rm e}^{-i\nu x}}.
\end{equation}
Throughout the paper and in numerical simulations, we use the direct expression for the dipole acceleration $\ddot{d}=\left\langle\psi\left|\hat{a}\right|\psi\right\rangle$~\cite{Haes11} although the dipole signal can as easily be used. We also use a Hanning window~\cite{MesPowSpec} over the time duration of the simulations to avoid spurious frequencies in the computation of the associated discrete Fourier transform~(\ref{eq:HHG_spectrum}) of finite time signals.

We now introduce the model we use in numerical simulations. We consider a two-center soft-Coulomb potential~\cite{Java88} where the electron dynamics is along a line that forms an angle $\theta$ with the molecular axis, as illustrated in Fig.~\ref{fig:Potential_illustration}. The potential has the form:
\begin{eqnarray}
	\mathcal{V}_{\theta}\left(x\right) & = & 
		- \frac{Z_{eff}}{\sqrt{x^{2} - R x \cos\theta + \frac{R^{2}}{4} + a^{2}}} \nonumber \\ &&
		- \frac{Z_{eff}}{\sqrt{x^{2} + R x \cos\theta + \frac{R^{2}}{4} + a^{2}}}, \label{eq:Potential}
\end{eqnarray}
where $Z_{eff}$ is the effective charge, $R$ the internuclear distance and $a$ the softening parameter. We consider two sets of parameters, $Z_{eff}=1$ and $R=2$ or $Z_{eff}=0.5$ and $R=0.5$, with the softening parameter set such that the field-free ionization potential is $I_{p}=1$ or $I_{p}=0.5$ respectively ($a\approx1.39$ and $a\approx1.33$). Each can be seen as rough approximations of the ${\rm H}_{2}^{+}$ molecular ion and the ${\rm H}_{2}$ molecule, respectively, and will be referred to as such in what follows. We introduce the angle $\theta$ so as to investigate the changes in the HHG spectrum as the polarization direction is varied and how the factorizations (see section~\ref{sec:HHG_factorization}) reproduce these changes.

Although the discussion is kept as general as possible, for the numerical simulations reported in this paper, we consider a linearly polarized laser field with a constant envelope
\begin{equation} \label{eq:Laser_field}
	\mathcal{E}\left(t\right) = \mathcal{E}_{0}\cos\left(\omega t\right),
\end{equation}
where $\mathcal{E}_{0}$ is the peak field amplitude and $\omega$ the laser frequency and all simulations are started at a zero of the field. For numerical integration of the TDSE, we use a second order pseudo-spectral split operator scheme where the kinetic part is treated in momentum space (using fast-Fourier transforms) and the potential part in configuration space~\cite{Band13}, initialized in the ground state. In all cases we use high resolution computations and we have checked the robustness of the reported results with parameters.

\section{Semi-classical wavefunction} \label{sec:SCWF}

We will consider a HHG scenario in which ionization is kept low. Combined with the long wavelengths we consider in this paper, ionization will be assumed as an adiabatic process. At each time $t_{0}$, the instantaneous ionization rate is taken as the one for a static electric field with amplitude $\mathcal{E}\left(t_{0}\right)$ (see section~\ref{sec:SCWF:Bound_state_and_ionization}). In this section, we introduce the semi-classical wavefunction (SCWF) which is used to model and analyze the electron dynamics following ionization and the associated HHG emission.

\begin{figure}
	\centering
		\includegraphics[width=\linewidth]{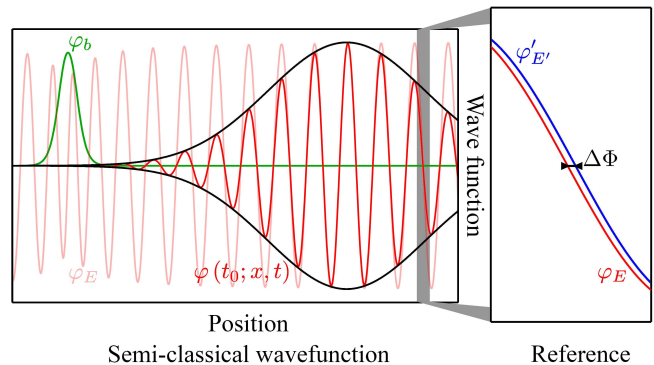}
	\caption{\label{fig:SCWF} (color online)
	Illustration of the semi-classical wavefunction (SCWF) $\varphi\left(t_{0};x,t\right)$ (full color, limited by the Gaussian envelope -- see text) and bound state $\varphi_{b}$. For comparison, we also display, in light shade, the continuum state $\varphi_{E}$ attached to the SCWF. The right part of the figure highlights the phase difference $\Delta\Phi\left(E,E^{\prime}\right)$ between the (continuum state) system $\varphi_{E}$ and reference $\varphi_{E^{\prime}}^{\prime}$.}
\end{figure}

We decompose the wavefunction between its bound and ionized parts
\begin{equation} \label{eq:Wavefunction_decomposition}
	\psi\left(x,t\right) = 
		\underbrace{\varphi_{b}\left(x\right) \alpha_{b}\left(t\right){\rm e}^{i\phi_{b}\left(t\right)}}_{bound} 
		+ \underbrace{\int^{t}{dt_{0}\ \varphi\left(t_{0};x,t\right)}}_{ionized},
\end{equation}
where the unspecified lower bound in the integral is set to the initial time for quantum simulations. We will discuss the bound part of the wave function in detail in section~\ref{sec:SCWF:Bound_state_and_ionization}. $\varphi\left(t_{0};x,t\right)$ corresponds to the subsequent dynamics of the part of the bound wavefunction ionized at time $t_{0}$. We show an illustration of the bound and ionized parts of the wavefunction in the SCWF approximation in Fig.~\ref{fig:SCWF}.

The strong field ionization process has drawn a lot of attention in the adiabatic regime and beyond from the seminal works of Keldysh~\cite{Keld65}, Perelomov-Popov-Terentev~\cite{Pere66} and Ammosov-Delone-Krainov~\cite{Ammo86} to TDSE numerical approaches~\cite{Baue99,Tong05}. Generally speaking, these theories predict the electron, after exiting the ionization barrier at time $t_{0}$, to exhibit a Gaussian distribution in momentum, generally centered around $0$ for linear polarization. Moving the momentum distribution picture into configuration space, we take the initial ionized part of the wavefunction as a Gaussian profile. For the subsequent dynamics, the SCWF approximation consists of two main hypotheses. 
(1)~We assume that the spatial profile remains Gaussian, with a time-dependent maximum $x\left(t_{0};t\right)$ and width $\sigma\left(t_{0};t\right)$. Following the maximum of the Gaussian then gives rise to a classical trajectory in phase space, with position and momentum $x\left(t_{0};t\right)$ and $p\left(t_{0};t\right)$, respectively.
(2)~We assume that the fast spatial variations of the SCWF can be described by the (single, field-free) continuum state with the corresponding momentum $p\left(t_{0};t\right)$ around $x\left(t_{0};t\right)$.
For simplicity, we label this continuum state $\left|\varphi_{E}\right\rangle$ with its energy $E\left(t_{0};t\right)$ following the electron dynamics. The motivation for using continuum states rather than the usual plane wave/Volkov states/Coulomb waves/\ldots\ is to account for the specific influence of the potential at hand~\cite{Smir09_2,Worn09,Lin10}, e.g., when the electron is close to the core region. Altogether, the SCWF model then yields:
\begin{equation} \label{eq:SCWF}
	\left|\varphi\left(t_{0};t\right)\right\rangle = 
		\alpha_{0}\left(t_{0}\right){\rm e}^{i\phi_{0}\left(t_{0}\right)}
		\frac{
			{\rm e}^{-\frac{\left(x-x\left(t_{0};t\right)\right)^{2}}{4\sigma^{2}\left(t_{0};t\right)}}
		}{
			\sqrt{\sigma\left(t_{0};t\right)}
		}
		\left|\varphi_{E\left(t_{0};t\right)}\right\rangle{\rm e}^{i\phi\left(t_{0};t\right)},
\end{equation}
where $\alpha_{0}$ and $\phi_{0}$ are related to the ionization yield and phase respectively (see section~\ref{sec:SCWF:Bound_state_and_ionization}), $\phi$ corresponds to the phase accumulated by the SCFW trajectory, while $\sigma$ accounts for the quantum spread of the wavefunction in the continuum (see section~\ref{sec:SCWF:Trajectory}). 
In the past, the use of semi-classical frozen Gaussians~\cite{Hell81,Kluk86,Kay94} have proven very useful in several fields of physical-chemistry with a more recent application to HHG~\cite{Zago12}.

\subsection{Electronic trajectory} \label{sec:SCWF:Trajectory}

In the simple case of a flat potential -- corresponding to the strong-field approximation (SFA) -- continuum states can be computed analytically and correspond to the so-called Volkov states~\cite{Popr14} leading to
$$
	\left|\varphi_{E\left(t_{0};t\right)}\right\rangle{\rm e}^{i\phi\left(t_{0};t\right)} = \frac{1}{\sqrt[4]{2\pi}}
		{\rm e}^{i \left(p\left(t_{0};t\right)x - \int_{t_{0}}^{t}{ds\ \frac{p^{2}\left(t_{0};s\right)}{2}}\right)},
$$
with $E\left(t_{0};t\right)=p^{2}\left(t_{0};t\right)/2$. In this case, the semi-classical trajectory is given by Hamilton's equations
$$
	d_{t} x\left(t_{0};t\right) = p\left(t_{0};\cdot\right) \ \ \ {\rm and} \ \ \ 
	d_{t} p\left(t_{0};t\right) = -\mathcal{E},
$$
and can for instance be found using the stationary phase approximation. We also note that the SCWF phase corresponds to the Hamiltonian action $\phi\left(t\right)=\int_{t_{0}}^{t}{ds\ p^{2}\left(t_{0};s\right)/2}$~\cite{Popr14}. In the SFA, the potential is flat so that its effect on the quantum dynamics is independent of the electron motion. In this context, the standard deviation can be derived from the free particle case giving
$$
	\sigma\left(t_{0};t\right) = \sqrt{\frac{4\sigma_{0}^{4}+\left(t-t_{0}\right)^{2}}{4\sigma_{0}^{2}}},
$$
where $\sigma_{0}=\sigma\left(t_{0};t_{0}\right)$ is therefore the initial standard deviation, immediately after the ionization step. Note that our definition of the SCWF naturally avoids the singularity of the standard deviation for $t\to t_{0}$ as is typically observed using the stationary phase approximation~\cite{Lewe94,Pugl15}. Numerically we find that in the SFA the SCWF approximation offers very good results compared to the full quantum dynamics.

Beyond the SFA, e.g., for long range potentials with a Coulomb-like tail, one can consider substituting Coulomb waves instead of the Volkov states in the previous equations and adapt the subsequent analysis accordingly. Generally speaking, for a given potential, the SCWF approximation consists of selecting an appropriate dynamics for the position, momentum, standard deviation, phase and energy. Irrespective of this specific choice, in what follows we will assume
\begin{equation} \label{eq:SCWF_trajectory}
	\dot{x} \approx \partial_{p}\mathcal{H}, \ \ \ 
	\dot{p} \approx -\partial_{x}\mathcal{H} \ \ \ {\rm and} \ \ \ 
	\dot{\phi} \approx -E,
\end{equation}
where $\mathcal{H}$ is the classical counterpart to the Hamiltonian operator $\hat{\mathcal{H}}$, and we use Hamilton's equations in place of the TDSE. The phase derivative equation ensures that, in the limit of infinite standard deviation $\sigma\to\infty$, the quantum dynamics for the continuum state $i\partial_{t}\left|\varphi_{E}\right\rangle{\rm e}^{i\phi}=\hat{\mathcal{H}}\left|\varphi_{E}\right\rangle{\rm e}^{i\phi}=E\left|\varphi_{E}\right\rangle{\rm e}^{i\phi}$ is satisfied. As we shall see, this approximation is very useful whenever the phase derivative is needed in theoretical investigations.

\subsection{Bound state and ionization} \label{sec:SCWF:Bound_state_and_ionization}

For the sake of simplicity, we assume the bound part of the wavefunction to be a single, field-free, eigenstate~\cite{Lewe94} -- in most cases the ground state -- which we denote $\varphi_{b}$ as illustrated in the left part of Fig.~\ref{fig:Potential_illustration}. We neglect laser induced spatial variations of the bound state and compute the complex ionization potential $I_{p}\left(\mathcal{E}\right)$ for a static electric field with amplitude $\mathcal{E}$, using a complex rotation~\cite{Rein82}. Then, the bound part dynamics of Eq.~(\ref{eq:Wavefunction_decomposition}) is given by
\begin{equation} \label{eq:Bound_state_dynamics}
	\alpha_{b}\left(t\right) =  
		{\rm e}^{-\int^{t}{ds\ \frac{\Gamma_{b}\left(\mathcal{E}\left(s\right)\right)}{2}}}, \ \ \ 
	\phi_{b}\left(t\right) = 
		-\int^{t}{ds\ E_{b}\left(\mathcal{E}\left(s\right)\right)},
\end{equation}
where $E_{b}$ is the Stark-shifted bound-state energy and $\Gamma_{b}$ the ionization rate with $I_{p}=-E_{b}+i\Gamma_{b}/2$.

From the bound state dynamics, we can now derive the initial, ionization step~$(i)$, condition for the ionized part of the wavefunction in Eq.~(\ref{eq:SCWF}). Indeed, given that bound and continuum eigenstates of the Hamiltonian operator form a generalized orthonormal basis, if we neglect recapture of previously ionized electrons, charge conservation imposes
$$
	\alpha_{b}^{2}\left(t\right) + \int^{t}{dt_{0}\ \alpha_{0}^{2}\left(t_{0}\right)} \approx 1,
$$
where we have neglected the effect of the Gaussian profile on the cross terms when computing the total charge. Taking the derivative of the previous equality and using Eq.~(\ref{eq:Bound_state_dynamics}) after a short calculation we get
\begin{equation} \label{eq:SCWF:Amplitude}
	\alpha_{0}\left(t_{0}\right) = 
		\sqrt{\Gamma_{b}\left(\mathcal{E}\left(t_{0}\right)\right)} 
		\exp\left(-\int^{t_{0}}{ds\ \frac{\Gamma_{b}\left(\mathcal{E}\left(s\right)\right)}{2}}\right).
\end{equation}
In the low ionization regime, when bound state depletion can be ignored, the amplitude coefficient simply becomes $\alpha_{0}\approx\sqrt{\Gamma_{b}}$. Finally, the adiabatic approximation applied to the ionization phase leads to
\begin{equation} \label{eq:SCWF:Ionization_phase}
	\phi_{0}\left(t_{0}\right) = -\int^{t_{0}}{ds\ E_{b}\left(\mathcal{E}\left(s\right)\right)} +
		\Delta\phi_{0}\left(\mathcal{E}\left(t_{0}\right)\right),
\end{equation}
where $\Delta\phi_{0}\left(\mathcal{E}\left(t_{0}\right)\right)$ is the phase accumulated only during the ionization process and the first term on the right-hand side reflects the synchronization of the ionized part of the wavefunction with the bound state phase at the instant of ionization.

\subsection{Dipole radiation signal} \label{sec:SCWF:Dipole_signal}

We now have all the key ingredients to express the dipole signal and, in turn, its associated HHG spectrum using the SCWF approximation for the wavefunction dynamics. From Eq.~(\ref{eq:Wavefunction_decomposition}), combined with the dipole definition~(\ref{eq:HHG_spectrum}), we identify three main contributions to the dipole signal. From these three, we ignore the contributions from the bound-bound and continuum-continuum state coupling (due to its low frequency spectrum, and its second-order importance compared to the bound-continuum component, respectively). We isolate the contributions from each initial ionization time and define the complex dipole acceleration element
\begin{equation} \label{eq:Complex_dipole}
	\ddot{d}\left(t_{0};t\right) = 
		\left\langle \varphi_{b}\left| \hat{a} \right|\varphi\left(t_{0};x,t\right)\right\rangle
		\alpha_{b}\left(t\right) {\rm e}^{-i\phi_{b}\left(t\right)},
\end{equation}
where the bound state amplitude and phase are given by Eqs.~(\ref{eq:Bound_state_dynamics}) and the total dipole acceleration is obtained by integrating over ionization times
$$
	\ddot{d}\left(t\right) = \int^{t}{dt_{0}\ \ddot{d}\left(t_{0};t\right)} + c.c.,
$$
with $c.c.$ the complex conjugate. From the definition of the SCWF~(\ref{eq:SCWF}), after appropriate factorization, one can isolate the three steps of the recollision model in the complex acceleration dipole
\begin{widetext}
\begin{equation} \label{eq:Complex_dipole:factorization}
	\ddot{d}\left(t_{0};t\right) = 
		\underbrace{
			\sqrt{\Gamma_{b}\left(\mathcal{E}\left(t_{0}\right)\right)}
			{\rm e}^{
				-\int^{t_{0}}{ds\ \Gamma_{b}\left(\mathcal{E}\left(s\right)\right)}
				+i\Delta\phi_{0}\left(\mathcal{E}\left(t_{0}\right)\right)
			}
		}_{(i)\ {\rm ionization}}
		\underbrace{
			\frac{
				{\rm e}^{i\left(
					\phi\left(t_{0};t\right) - \int_{t_{0}}^{t}{ds\ I_{p}^{*}\left(\mathcal{E}\left(s\right)\right)}
				\right)}
			}{\sqrt{\sigma\left(t_{0};t\right)}}
		}_{(ii)\ {\rm propagation}}
		\underbrace{
			\left\langle \varphi_{b} \bigg| \hat{a} \bigg|
				{\rm e}^{-\frac{\left(x-x\left(t_{0};t\right)\right)^{2}}{4\sigma^{2}\left(t_{0};t\right)}}
				\varphi_{E\left(t_{0};t\right)} \right\rangle
		}_{(iii)\ {\rm rescattering}},
\end{equation}
\end{widetext}
where ``$I_{p}^{*}$'' is the complex conjugate of the ionization potential. The first step,$(i)$ ionization, has been set by hand in the model through the adiabatic approximation and there is therefore little surprise to find it here. On the other hand, the clear separation between the second,$(ii)$ propagation, and the third,$(iii)$ rescattering, was not predetermined and is a direct consequence of the SCWF model. 

\section{Factorization of the high harmonic generation spectrum} \label{sec:HHG_factorization}

The factorization~(\ref{eq:Complex_dipole:factorization}) makes clear that the complex dipole signal associated with a given SCWF is the product of the ionization, propagation and recollision cross sections. Note though that this factorization is expressed here in the time domain. In this section we investigate how the factorization maps to the frequency domain and, more interestingly for our purpose, to HHG spectra. First we investigate the factorization when the propagation part is described with a a reference system that only shares generic features with the system at hand, for instance a long-range Coulomb tail away from the core, as in the QRS formalism~\cite{Lin10,Le08,Le13} (section~\ref{sec:HHG_factorization:Reference_system}). While the use of such a reference provides very good results for HHG spectrum predictions, it sheds little light on the propagation step~$(ii)$ which is treated as a black box. Alternatively, this question can be investigated with the SCWF perspective taking advantage of the trajectory component of the model~(section~\ref{sec:HHG_factorization:Direct_factorization}). More specifically, it allows us to disentangle the individual contributions from all the trajectories that contribute to the HHG spectrum and enables us to compare the relative contributions $(i)-(iii)$.

For consistency with our choice of computing HHG spectra using the acceleration form of the dipole signal, in what follows we discuss the factorization using the acceleration scattering cross-section $\left\langle\varphi_{b}\left|\hat{a}\right|\varphi_{E}\right\rangle=-\left\langle\varphi_{b}\left|\nabla\mathcal{V}\right|\varphi_{E}\right\rangle$. A similar analysis can be carried out using the dipole form and its associated scattering cross-section $\left\langle\varphi_{b}\left|\hat{x}\right|\varphi_{E}\right\rangle$. The equivalence between the two factorization forms will be discussed in the end of section~\ref{sec:HHG_factorization:Reference_system}.

\subsection{Using a reference system} \label{sec:HHG_factorization:Reference_system} 

The central idea behind using a reference system is to approximate the propagation term $(ii)$ as a black box by including some of the effects of the potential at hand. To be a good candidate, the reference system should be easy to compute and/or common to a wide range of parameters under investigation where the reference is computed once and then reused throughout the parameter analysis. For our illustration of molecular models of potential~(\ref{eq:Potential}) we investigate the dependance of the HHG spectrum with the polarization angle $R_{\rm HHG}\left(\theta;\nu\right)$. Similarly to the QRS formalism we define the reference spectrum $R_{\rm HHG}^{\prime}\left(\nu\right)$ -- more generally we will label all data associated with the reference system with primes -- with identical (field-free) ionization potential~\cite{note_Ip} and similar potential shape away from the core region. In our case, such a reference can be taken to be the system at a given angle, e.g., $R_{\rm HHG}^{\prime}\left(\nu\right)=R_{\rm HHG}\left(\pi/2;\nu\right)$ where the potential becomes the one for a SAE atomic target as illustrated in Fig.~\ref{fig:Potential_illustration}. Then, from the definition of the HHG spectrum~(\ref{eq:HHG_spectrum}) and using the linearity of the Fourier transform we apply the SWCF approximation to the dipole accelerations~(\ref{eq:Complex_dipole:factorization}) and get
\begin{equation} \label{eq:HHG_spectrum:Reference_atom}
	R_{\rm HHG}\left(\theta;\nu\right) = \int{dt_{0}\ 
			\mathcal{F}\left[ 
				\ddot{d}^{\prime}\left(t_{0};t\right)\frac{\ddot{d}\left(\theta,t_{0};t\right)}{\ddot{d}^{\prime}\left(t_{0};t\right)}
			\right]\left(\nu\right) + c.c.
		},
\end{equation}
with $c.c.=\mathcal{F}\left[\ddot{d}^{*}\left(\theta,t_{0};t\right)\right]$. Intuitively, we see that the key element of the QRS factorization -- e.g., Eqs.~(1) and~(4) of Ref.~\cite{Le13} -- consists of moving the relative dipole acceleration $\ddot{d}/\ddot{d}^{\prime}$ outside of the Fourier transform as a global multiplicative factor. In what follows we investigate the theoretical grounds for doing so.

As discussed previously, we consider HHG in the low ionization regime where bound state depletion can be neglected such that the ionization part~$(i)$ of the ratio $\ddot{d}/\ddot{d}^{\prime}$ simplifies to
$$
	\frac{(i)}{(i)^{\prime}} \approx 
		\sqrt{\frac{\Gamma_{b}\left(\theta,t_{0}\right)}{\Gamma_{b}^{\prime}\left(t_{0}\right)}}
		{\rm e}^{i\Delta\Phi_{0}\left(\theta,\mathcal{E}\left(t_{0}\right)\right)},
$$
where $\Delta\Phi_{0}\left(\theta,\mathcal{E}\right)=\Delta\phi_{0}\left(\theta,\mathcal{E}\right)-\Delta\phi_{0}^{\prime}\left(\mathcal{E}\right)$ is the ionization phase difference with the reference system. Numerical computations of ionization rates show a very generic shape for systems with comparable ionization potential such that, at the leading order, $\sqrt{\Gamma_{b}\left(\theta,t_{0}\right)/\Gamma_{b}^{\prime}\left(t_{0}\right)}\approx\Gamma\left(\theta\right)$, irrespective of the ionization time $t_{0}$. For higher laser intensities, where the bound state depopulation effects can be neglected over \emph{one} laser cycle but not for the \emph{full} duration of the pulse, using the argument that similar dipole signals are produced every laser period, the previous equation can be modified taking into account the ionization yield over a laser cycle. Because of the integration over the laser cycle, the ionization factor is independent of the ionization time $t_{0}$ within a cycle. As a consequence, integrating over the laser pulse duration, the overall ionization ratio is also independent of $t_{0}$ and $\left|(i)/(i)^{\prime}\right|\approx\Gamma\left(\mathcal{E}_{0};\theta\right)$.

We now turn to the propagation part~$(ii)$ of the dipole acceleration ratio. This term describes the ionized electron dynamics in the continuum, i.e., mostly when the electron is far away from the core. Beyond the SFA picture, using a reference system with similar potential shape offers a much better description of this electron dynamics away from the core. We illustrate this point in the right part of Fig.~\ref{fig:SCWF} where only a zoom allows to differentiate between the system and reference continuum states away from the core and this difference is associated with a phase shift ($\Delta\Phi$). In the previous paragraph, we have ruled out effects of the bound state depopulation such that the only possible source of difference between the system and reference comes from their respective bound state energy Stark-shift. We assume that the Stark shift of the system and its reference are the same to leading order (which is a good approximation at moderate intensity) and would therefore cancel in the ratio $(ii)/(ii)^{\prime}$. Now looking at the overall dipole acceleration~(\ref{eq:Complex_dipole:factorization}) we notice that the phase term in $(ii)$ is the only fast oscillating factor, compared to all the other terms which vary on the time scale imposed by the laser frequency $\omega$. The instantaneous HHG frequency, defined as the time derivative of the total phase, is then:
$$
	\left|\nu\left(t_{0},t\right)\right| 
		\approx \left|\dot{\phi}\left(t_{0},t\right) - \Re\left(I_{p}\left(\mathcal{E}\left(t\right)\right)\right)\right| 
		\approx E\left(t_{0},t\right) + I_{p},
$$
where $\Re$ denotes the real part and we have used Eq.~(\ref{eq:SCWF_trajectory}) for the phase derivative. This instantaneous frequency will be discussed in further detail in section~\ref{sec:HHG_factorization:Direct_factorization}, for now it provides a link between the HHG frequency and electronic energy.

\begin{figure}
	\centering
		\includegraphics[width=\linewidth]{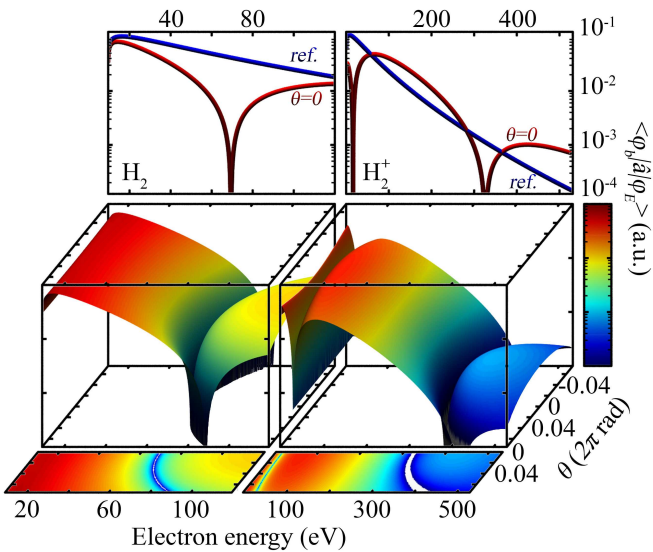}
	\caption{\label{fig:Scattering_cross_section} (color online)
	Polarization angle resolved acceleration scattering cross-sections with the ground state $\left\langle\varphi_{b}\left|\hat{a}\right|\varphi_ {E}\right\rangle$ for the ${\rm H}_{2}$ molecule (left panels) and ${\rm H}_{2}^{+}$ molecular ion (right). For clarity, the lower panels show the projection of the scattering cross-section over positive angles $\theta$ and the upper panels compares the cross-section curves for the reference system ($\theta=\pi/2$) and potential with parallel molecular and polarization directions ($\theta=0$, see labels).}
\end{figure}

For atomic and small molecular systems the bound part of the wavefunction $\varphi_{b}^{(\prime)}$ is localized in space and we define $\chi_{b}$ as the characteristic function over this region, i.e., $\chi_{b}\left(x\right)=0$ where $\varphi_{b}\approx0$ and $1$ elsewhere. Without loss of generality we use the same characteristic function for both the system and reference. This could for instance be achieved by increasing the respective characteristic domains to make them match. As is illustrated in Fig.~\ref{fig:SCWF}, the width of the ionized part of the wavefunction is typically much larger than that of the bound part. Intuitively, this can be understood by the fact that right after ionization the electron is usually localized in momentum space. Moving this picture in position initializes the ionized part of the wavefunction with a relatively large width $\sigma^{(\prime)}_{0}$ and is further amplified through quantum spread in the continuum. As a consequence we have
$$
	{\rm e}^{-\frac{\left(x-x\left(\theta,t_{0};t\right)\right)^{2}}{4\sigma^{2}\left(\theta,t_{0};t\right)}}
		\chi_{b}\left(x\right) \approx\frac{
			\int{dy\  {\rm e}^{-\frac{\left(y-x\left(\theta,t_{0};t\right)\right)^{2}}{4\sigma^{2}\left(\theta,t_{0};t\right)}}
				\chi_{b}\left(y\right)}
		}{
			\int{dy\ \chi_{b}\left(y\right)}
		}\chi_{b}\left(x\right),
$$
which approximates the Gaussian envelope with its mean value over the characteristic function $\chi_{b}$ domain, and thus
\begin{eqnarray}
	&& \left\langle \varphi_{b}\left(\theta\right) \bigg| \hat{a}\left(\theta\right) \bigg|
		{\rm e}^{-\frac{\left(x-x\left(\theta,t_{0};t\right)\right)^{2}}{4\sigma^{2}\left(\theta,t_{0};t\right)}}
		\varphi_{E\left(\theta,t_{0};t\right)}\right\rangle \approx \nonumber \\ && \quad
	\frac{
			\int{dy\  {\rm e}^{-\frac{\left(y-x\left(\theta,t_{0};t\right)\right)^{2}}{4\sigma^{2}\left(\theta,t_{0};t\right)}}
				\chi_{b}\left(y\right)}
		}{
			\int{dy\ \chi_{b}\left(y\right)}
		} \left\langle \varphi_{b}\left(\theta\right) \left| \hat{a}\left(\theta\right) \right|
		\varphi_{E\left(\theta,t_{0};t\right)} \right\rangle. \label{eq:Dipole_acceleration_element}
\end{eqnarray}
A similar approximation can be written for the reference system.  Again, because of the similarity between the potential with the reference model we expect $x\left(\theta,t_{0};t\right) \approx x^{\prime}\left(t_{0};t\right)$ in the region away from the core such that the prefactors to the bound-continuum acceleration dipole element in the previous equation cancel in the rescattering part of the dipole acceleration ratio
$$
	\frac{(iii)}{(iii)^{\prime}} \approx \frac{
		\left\langle \varphi_{b}\left(\theta\right) \left| \hat{a}\left(\theta\right) \right| 
			\varphi_{E\left(\theta,t_{0};t\right)}\left(\theta\right)\right\rangle
	}{
		\left\langle \varphi_{b}^{\prime} \left| \hat{a}^{\prime} \right| \varphi_{E^{\prime}\left(t_{0};t\right)}\right\rangle
	}.
$$
Already we see the central role played by the scattering states in the HHG spectrum, as emphasized by the QRS approximation~\cite{Lin10,Le08,Le13} as compared to the SFA where plane waves are used. In Fig.~\ref{fig:Scattering_cross_section} we display the angle resolved cross-section for the molecular models~(\ref{eq:Potential}) we consider here. For both molecular models we notice that the two center interference generates a singularity in the scattering cross-section whose position in energy depends on the angle $\theta$. Note, though, that when the laser is perpendicular to the molecular axis ($\theta=\pi/2$, see upper part of the Fig.~\ref{fig:Scattering_cross_section}), the singularity disappears, which makes this a good candidate as a reference since it avoids dividing by zero in the previous equation.

We put together all the simplifications discussed above for components $(i)-(iii)$ of the dipole acceleration ratio and combine it with the HHG spectrum~(\ref{eq:HHG_spectrum:Reference_atom}) which becomes
\begin{widetext}
\begin{equation} \label{eq:HHG_spectrum:Reference_atom:Convolution}
	R_{\rm HHG}\left(\theta;\nu\right) \approx \Gamma\left(\mathcal{E}_{0};\theta\right) \int{dt_{0}\ 
			{\rm e}^{i\Delta\Phi_{0}\left(\theta,\mathcal{E}\left(t_{0}\right)\right)} \left(
				\mathcal{F}\left[\ddot{d}^{\prime}\left(t_{0};t\right)\right] * \frac{1}{\sqrt{2\pi}}
				\mathcal{F}\left[\frac{
					\left\langle\varphi_{b}\left(\theta\right) \left| \hat{a} \right|
						\varphi_{E\left(\theta,t_{0};t\right)}\left(\theta\right)\right\rangle
				}{
					\left\langle\varphi_{b}^{\prime} \left| \hat{a}^{\prime} \right|
						\varphi_{E^{\prime}\left(t_{0};t\right)}\right\rangle
				}\right]
			\right)\left(\nu\right) 
		} + c.c.,
\end{equation}
\end{widetext}
given the convolution property of the Fourier transform of a product ($\mathcal{F}\left[fg\right]=\mathcal{F}\left[f\right]*\mathcal{F}\left[g\right]/\sqrt{2\pi}$). As discussed previously, the energy $E^{\left(\prime\right)}\left(t_{0};t\right)$ evolves on the time scale of the electron dynamics, which is very slow compared to the overall dipole variation associated with the total phase in propagation term~$(ii)$. As a consequence, this slow variation is recovered in the scattering cross-section ratio
\begin{equation} \label{eq:HHG_spectrum:Scattering_cross-section_ratio}
	\mathcal{F}\left[\frac{(iii)}{(iii)^{\prime}}\right]\left(\nu\right) \propto
		\delta\left(\nu-\omega\right) \approx \delta\left(\nu\right),
\end{equation}
where $\delta$ is the Dirac delta distribution, and given that $\omega\ll\nu$ for high harmonic orders. Following the time scale separation of the dipole signal phase discussed previously, the proportionality coefficient in the Fourier transform of Eq.~(\ref{eq:HHG_spectrum:Scattering_cross-section_ratio}) is obtained using the instantaneous frequency approximation and reads
$$
	\frac{
		\left\langle \varphi_{b}\left(\theta\right) \left| \hat{a}\left(\theta\right) \right| 
			\varphi_{E\left(t_{0};t\right)}\left(\theta\right)\right\rangle
	}{
		\left\langle \varphi_{b}^{\prime} \left| \hat{a}^{\prime} \right| \varphi_{E^{\prime}\left(t_{0};t\right)}\right\rangle
	} \ \ \ {\rm with} \ \ \ E=\left|\nu\right|-I_{p}
$$
in Eq.~(\ref{eq:HHG_spectrum:Reference_atom:Convolution}). Since the reference system dynamics away from the core is assumed to reproduce the one for the system at hand we further consider $E\approx E^{\prime}$. The last term we are left to deal with in the HHG spectrum factorization is the ionization phase difference $\Delta\Phi_{0}$. Most ionization models attribute similar ionization effects to potentials with identical field-free ionization potentials and we therefore ignore this additional phase altogether $\Delta\Phi_{0}\approx0$. In the end, we arrive at the HHG spectrum factorization from the reference system
\begin{equation} \label{eq:HHG_spectrum:Reference_atom:Factorization}
	R_{\rm HHG}\left(\theta,\nu\right) \approx \Gamma\left(\mathcal{E}_{0};\theta\right) R_{\rm HHG}^{\prime}\left(\nu\right)
		\frac{\left\langle\varphi_{b}\left|\hat{a}\right|\varphi_{E}\right\rangle\left(\theta\right)}{
		\left\langle\varphi_{b}^{\prime}\left|\hat{a}^{\prime}\right|\varphi_{E}^{\prime}\right\rangle},
\end{equation}
given that $R_{\rm HHG}^{\prime}=\int{dt_{0}\ \mathcal{F}\left[\ddot{d}^{\prime}\left(t_{0};t\right)\right]}+c.c.$. As emphasized in the QRS formulation~\cite{Lin10}, the dipole acceleration element ratio $\left\langle\varphi_{b}\left|\hat{a}\right|\varphi_{E}\right\rangle/\left\langle\varphi_{b}^{\prime}\left|\hat{a}^{\prime}\right|\varphi_{E}^{\prime}\right\rangle$ contains phase information due to the phase difference of field-free continuum eigenstates (denoted $\Delta\Phi$ on the right most part of Fig.~\ref{fig:Potential_illustration}). We finish by noticing that one can also substitute the dipole element ratio 
$$
	\frac{\left\langle\varphi_{b}\left|\hat{a}\right|\varphi_{E}\right\rangle}{
		\left\langle\varphi_{b}^{\prime}\left|\hat{a}^{\prime}\right|\varphi_{E}^{\prime}\right\rangle} \approx
	\frac{\left\langle\varphi_{b}\left|\hat{x}\right|\varphi_{E}\right\rangle}{
		\left\langle\varphi_{b}^{\prime}\left|\hat{x}^{\prime}\right|\varphi_{E}^{\prime}\right\rangle},
$$
in Eq.~(\ref{eq:HHG_spectrum:Reference_atom:Factorization}), e.g., using the approximation $\left\langle\varphi_{b}\left|\hat{a}\right|\varphi_{E}\right\rangle\approx-E^{2}\left\langle\varphi_{b}\left|\hat{x}\right|\varphi_{E}\right\rangle$, and that this corresponds to the standard formulation of the QRS factorization~\cite{Lin10}.

\begin{figure}
	\centering
		\includegraphics[width=\linewidth]{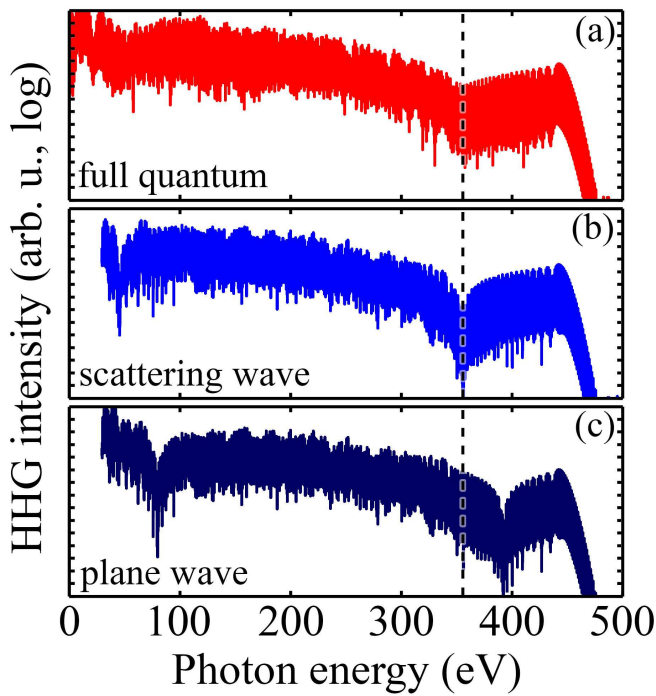}
	\caption{\label{fig:HHG_reference_system} (color online)
	HHG spectrum for the ${\rm H}_{2}^{+}$ molecular ion model of potential~(\ref{eq:Potential}) with parallel molecular and polarization directions ($\theta=0$). For the simulations, we integrate the TDSE over 10 laser cycles with $3\times10^{14}~{\rm W}\cdot{\rm cm}^{-2}$ and $2150~{\rm nm}$ laser intensity and wavelength, respectively. From up to down, we compare spectra obtained from (a) the full quantum dynamics with Eq.~(\ref{eq:HHG_spectrum}) to (b-c) the reference system factorization~(\ref{eq:HHG_spectrum:Reference_atom:Factorization}) using scattering and plane waves respectively~\cite{note_Ip}. For indication, on each panel, the vertical dashed line labels the HHG energy associated with the singularity in the acceleration scattering cross section as seen in the upper right panel of Fig.~\ref{fig:Scattering_cross_section}.}
\end{figure}

To conclude this section, in Fig.~\ref{fig:HHG_reference_system} we compare the spectrum factorization using a reference system~(\ref{eq:HHG_spectrum:Reference_atom:Factorization}) with the result of a full quantum simulation -- panels (b) and (a) respectively. As demonstrated in QRS analyses~\cite{Lin10}, such a factorization offers a very good approximation of the actual result as the two spectra are very similar. In particular, we notice that both spectra exhibit a local minimum around $360~{\rm eV}$, labeled with vertical dashed lines, which is associated with a singularity in the scattering cross section (see upper right panel of Fig.~\ref{fig:Scattering_cross_section}). On the other hand, looking at the lower panel of Fig.~\ref{fig:HHG_reference_system} (c) we see that using the plane wave scattering cross-section in the spectrum factorization yields poor results (the local minimum is shifted by about $40~{\rm eV}$. This discrepancy illustrates the crucial importance of using continuum states for the system at hand, rather than plane waves (e.g., Volkov states), in the SCWF~(\ref{eq:SCWF}).

\subsection{Direct factorization} \label{sec:HHG_factorization:Direct_factorization}

As illustrated in Fig.~\ref{fig:HHG_reference_system}, and more generally discussed in QRS analyses~\cite{Lin10,Le08,Le13}, the use of  a reference system is interesting in that it offers very good, quantitatively comparable, results compared with the full quantum simulations. For appropriately chosen systems, such as a scaled hydrogen atom, the computation of the reference HHG spectrum can be evaluated numerically relatively cheaply with modern technology. Yet, beyond the computational point of view, from the theoretical perspective, one of the drawbacks of using a reference system is that it treats the propagation step~$(ii)$ as a black box from which little physical insight is gained. On the other hand, insightful electron trajectory pictures have been developed for the interpretation of HHG spectra, for instance the well-known short and long trajectories with linearly polarized lasers~\cite{Gaar08,Mair03}. In this section we connect the trajectory component of the SCWF to the propagation component~$(ii)$ of the spectrum which leads us to a direct factorization of the HHG spectrum. In particular, this analysis allows to separate the contributions from each such trajectory and to compare the relative importance of the three steps $(i)-(iii)$ to the spectrum.

In this context, the analysis is the same irrespective of the polarization angle and, for the sake of simplicity, in what follows we omit the $\theta$ parameter dependence in equations when there is no confusion possible. We start again from the SCWF approximation in which contributions to the harmonic spectrum are separated by ionization time. We define the element
\begin{equation} \label{eq:HHG_spectrum_element}
	R_{\rm HHG}\left(t_{0};\nu\right) = \mathcal{F}\left[\ddot{d}\left(t_{0};t\right)\right]\left(\nu\right),
\end{equation}
such that
$$
	R_{\rm HHG}\left(\nu\right) = \int{dt_{0}\ R_{\rm HHG}\left(t_{0};\nu\right) + c.c.}.
$$

For typical atomic and small molecular systems, the bound part of the wavefunction is localized in a well defined part of space, which we denoted with the characteristic function $\chi_{b}$ in the previous section. In comparison, the ionized electron dynamics extends over much larger excursion distances, as illustrated in Fig.~\ref{fig:SCWF}. As a consequence, the trajectory component of the SCWF model allows for the definition of a recollision time $t_{r}$ when the electron returns to the core (or the time of closest return depending on the chosen model), if any. In our case of potential~(\ref{eq:Potential}), the recollision time for a given trajectory is defined by the implicit equation $x\left(t_{0};t_{r}\right)=0$. Then, considering a linearization of the trajectory around this recollision time, combined with the comparatively large Gaussian width of the SCWF, the spatial averaging~(\ref{eq:Dipole_acceleration_element}) can be expressed in the temporal domain 
$$
	(iii) \approx {\rm e}^{-\frac{\left(t-t_{r}\right)^{2}}{2\tilde{\sigma}^{2}}}
		\left\langle\varphi_{b}\left|\hat{a}\right|\varphi_{E\left(t_{0};t\right)}\right\rangle,
$$
for some standard deviation $\tilde{\sigma}$ related to the parameters of the problem at hand.

Looking at the terms composing the complex dipole acceleration element~(\ref{eq:Complex_dipole:factorization}) we notice a clear separation of time scales between the different terms. On the one hand, the phase coefficient in propagation~$(ii)$, $\phi\left(t_{0};t\right)+\int_{t_{0}}^{t}{ds\ E_{b}\left(\mathcal{E}\left(s\right)\right)}$, exhibits a rapid variation in time. As introduced in the previous section we define its time derivative,
\begin{equation} \label{eq:Instantaneous_frequency}
	\nu\left(t_{0};t\right) = \dot{\phi}\left(t_{0};t\right) + E_{b}\left(\mathcal{E}\left(t\right)\right) \approx
		-E\left(t_{0};t\right) - I_{p}
\end{equation}
using the trajectory phase derivative approximation~(\ref{eq:SCWF_trajectory}) and neglecting any Stark shift. On the other hand, the term $\nu\left(t_{0};t\right)$ like the other time-dependent coefficients in Eq.~(\ref{eq:Complex_dipole:factorization}) evolves with the characteristic time scale of the SWCF electron dynamics, i.e., very slowly -- typically with the frequency of the driving laser $\omega$. We then consider a linearization of the phase around the recollision time which allows for a computation of the HHG spectrum element~(\ref{eq:HHG_spectrum_element})
\begin{equation}
	R_{\rm HHG}\left(t_{0};\nu\right) \approx (i) (ii) (iii) \Big|_{t=t_{r}}
		\tilde{\sigma} {\rm e}^{-\frac{\left(\nu-\nu\left(t_{0};t_{r}\right)\right)^{2}\tilde{\sigma}^{2}}{2}-i\nu t_{r}},
\end{equation}
where the second and third factors of Eq.~(\ref{eq:Complex_dipole:factorization}) are evaluated at the recollision time. In the limit of large $\tilde{\sigma}$ we notice that
$$
	\tilde{\sigma} {\rm e}^{-\frac{\left(\nu-\nu\left(t_{0};t_{r}\right)\right)^{2}\tilde{\sigma}^{2}}{2}} 
		\xrightarrow[\tilde{\sigma}\to\infty]{} \sqrt{2\pi} \delta\left(\nu-\nu\left(t_{0};t_{r}\right)\right),
$$
the Dirac delta distribution. Taking this limit when summing over ionization times, we see that the overall HHG spectrum adds up to the coherent superposition of the contributions from all ionization times leading to the same recollision frequency
\begin{equation} \label{eq:HHG_spectrum:Direct_factorization}
	R_{\rm HHG}\left(\nu\right) = \sum_{t_{0}^{\prime}}{ \ddot{d}\left(t_{0}^{\prime};t_{r}\right)}
		{\rm e}^{-i\nu t_{r}}\ {\rm with}\ t_{0}^{\prime} \ {\rm s.t.}\ \nu\left(t_{0}^{\prime};t_{r}\right)=\nu.
\end{equation}

From the previous equation~(\ref{eq:HHG_spectrum:Direct_factorization}) we recover the direct link between semi-classical trajectories of the SCWF and harmonics in the HHG spectrum. For example, in the SFA and linear polarization, the sum over ionization times $t_{0}^{\prime}$ corresponds to finding the short, long and possible multiple recollision trajectories leading to a given harmonic frequency. Breaking down the factors in the individual contributions for a given trajectory, we get
\begin{equation}
	\left|R_{\rm HHG} \left(t_{0}^{\prime};\nu\left(t_{0}^{\prime};t_{r}\right)\right)\right| \propto
		\sqrt{\frac{
			\Gamma_{b}\left(\mathcal{E}\left(t_{0}^{\prime}\right)\right)
		}{
			\sigma\left(t_{0}^{\prime};t_{r}\right)
		}} \left|
			\left\langle \varphi_{b} \left| \hat{a} \right| \varphi_{E} \right\rangle
		\right|,
\end{equation}
where we have neglected bound state depopulation and recall $E=\left|\nu\left(t_{0};t_{r}\right)\right|-I_{p}$~(\ref{eq:Instantaneous_frequency}). Furthermore, in the previous equation we remark that the variations of the $\sigma$ factor -- associated with quantum spread in the propagation term -- are much slower than that of the ionization and rescattering factors. This is made obvious in the SFA where $\sigma\left(t_{0}^{\prime};t_{r}\right)\approx\left|t_{r}-t_{0}^{\prime}\right|$ in the limit of large propagation times. We see that the linear dependence is negligible compared to the typical exponential variations over several orders of magnitude that normally occur in both the ionization and scattering cross-section (as can be seen in Fig.~\ref{fig:Scattering_cross_section} for the latter). From this perspective, the variations of the propagation factor can be neglected and the trajectory contribution is reduced to
\begin{equation} \label{eq:HHG_spectrum:Trajectory_contribution}
	\left| R_{\rm HHG}\left(t_{0}^{\prime};\nu\left(t_{0}^{\prime};t_{r}\right)\right) \right| \propto
		\sqrt{\Gamma_{b}\left(\mathcal{E}\left(t_{0}^{\prime}\right)\right)} 
		\left| \left\langle\varphi_{b}\left|\hat{a}\right|\varphi_{E}\right\rangle \right|.
\end{equation}

\begin{figure}
	\centering
		\includegraphics[width=\linewidth]{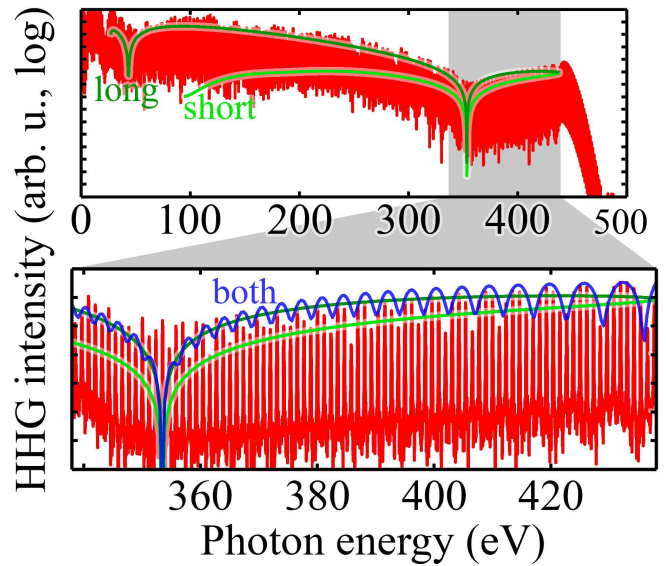}
	\caption{\label{fig:Spectrum_decomposition} (color online)
	Comparison between the full quantum HHG spectrum with the prediction~(\ref{eq:HHG_spectrum:Trajectory_contribution}) for short and long (upper panel, see labels) trajectories and the coherent superposition of both (lower) from Eq.~(\ref{eq:HHG_spectrum:Direct_factorization}). The proportionality coefficient in~(\ref{eq:HHG_spectrum:Trajectory_contribution}) has been chosen such as to get best match with the full HHG computation. As illustrated in the figure, the lower panel focuses on the part of the spectrum between $2.4Up+I_{p}$ and $3.17Up+I_{p}$ where only one short and one long trajectory (no multiple recollision) contribute to the spectrum in the SFA. The system (${\rm H}_{2}^{+}$), laser and computation parameters are the same as of Fig.~\ref{fig:HHG_reference_system}.}
\end{figure}

For simplicity, we use the SFA to compute the SCWF trajectory component -- although as discussed in section~\ref{sec:SCWF:Trajectory}, a more refined model including the effects of the potential could also be considered. In the upper panel of Fig.~\ref{fig:Spectrum_decomposition} we compare the contributions of short and long trajectories (see labels on the figure) using prediction~(\ref{eq:HHG_spectrum:Trajectory_contribution}) with the HHG spectrum of a full quantum simulation. In our configuration, we see that the long trajectory contribution qualitatively reproduces the overall shape of the full HHG spectrum and dominates the short trajectory component. This can be easily understood with the fact that long trajectories are born around the maxima of the electric field -- therefore with the higher ionization rate in the adiabatic approximation -- while short trajectories are initiated later on, when the instantaneous field is weaker. More generally note that, for a given harmonic energy, in the factorization framework we have developed here, both short and long (and multiple recollision) trajectories share the same rescattering cross section. As a consequence, the difference in their respective contribution amplitudes can only come from the ionization~$(i)$ and propagation~$(ii)$ factors.

Since more than one trajectory contributes to the HHG spectrum, following Eq.~(\ref{eq:HHG_spectrum:Direct_factorization}), they should be added coherently, that is including their respective phases. We display such a coherent superposition of both short and long trajectories, where their respective amplitudes are computed with Eq.~(\ref{eq:HHG_spectrum:Trajectory_contribution}), in the lower part of Fig.~\ref{fig:Spectrum_decomposition}. In the panel we focus on the harmonics between $2.4Up+I_{p}$ and $3.17Up+I_{p}$ where only one short and one long trajectory (no multiple recollision) contribute to the HHG spectrum according to the SFA, and $Up=\mathcal{E}_{0}^{2}/4\omega^2$ is the ponderomotive energy. When compared to the full quantum spectrum, we see that the coherent superposition of both short and long trajectories reproduces very well the oscillation pattern observed in the spectrum. We take it as a further proof of the relevance of the SCWF analysis and predictions of Eqs.~(\ref{eq:HHG_spectrum:Direct_factorization}) and~(\ref{eq:HHG_spectrum:Trajectory_contribution}).

Compared to the reference system version, the direct factorization bypasses the computation of a quantum HHG spectrum altogether. Beyond this kind of computational considerations, and more interestingly, the direct factorization offers an intuitive interpretation of spectra in terms of electron trajectories and allows to disentangle their respective contributions to the spectrum. This separation is essential in experimental measurements and models including macroscopic propagation of the HHG field where typical phase matching conditions restrict the contribution to a given harmonics from (at most) a single identified trajectory~\cite{Gaar08,Anto96,Zair08,Mair03}. In this context the individual spectra of isolated systems, as provided with the reference system factorization, are quite different from the macroscopic counterpart as it lacks the filtering imposed on trajectories that are not phase matched.

\section{Conclusions and perspectives} \label{sec:Conclusion_and_perspectives}

To summarize, we have introduced the semi-classical wavefunction (SCWF) approximation which combines the wave/particle picture of the electron dynamics: It is supported by a semi-classical trajectory while incorporating a spatially delocalized extension of the wavefunction. This intuitive framework, applied to high-harmonic generation (HHG) allows the factorization of the spectrum as the product of the ionization~$(i)$, propagation~$(ii)$ and rescattering~$(iii)$ cross-sections in energy (frequency) space. The propagation component can be described with a reference system~(\ref{eq:HHG_spectrum:Reference_atom:Factorization}) as in the quantitative rescattering (QRS) formalism~\cite{Lin10}. Alternatively, the factorization can be performed directly using the trajectory perspective of the SCWF~(\ref{eq:HHG_spectrum:Direct_factorization}) and~(\ref{eq:HHG_spectrum:Trajectory_contribution}).

\begin{figure*}
	\centering
		\includegraphics[width=\linewidth]{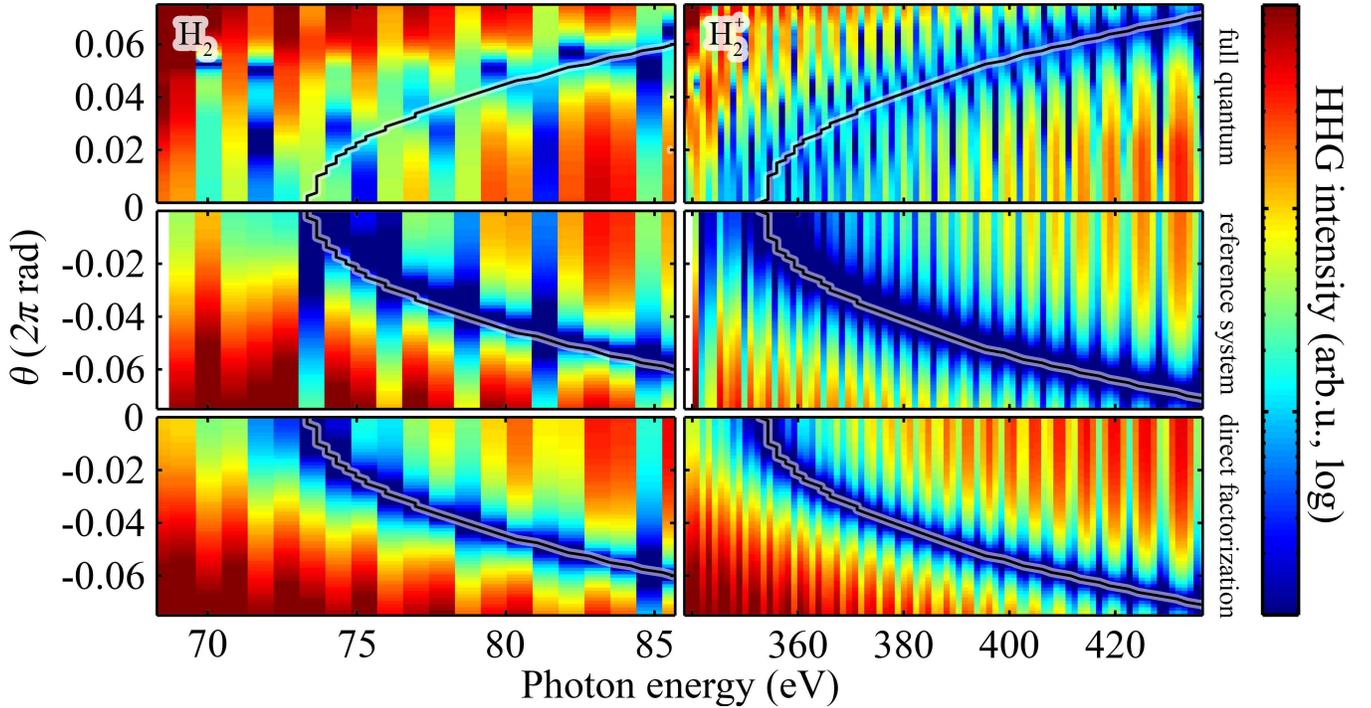}
	\caption{\label{fig:Angle_resolved_spectrum} (color online)
	Envelope of the HHG spectrum (obtained by selecting odd harmonics only) as the polarization direction angle $\theta$ is varied for the full quantum system (upper panel), using a reference system in factorization~(\ref{eq:HHG_spectrum:Reference_atom:Factorization})~\cite{note_Ip} (middle) and the direct factorization~(\ref{eq:HHG_spectrum:Direct_factorization}) and~(\ref{eq:HHG_spectrum:Trajectory_contribution}) with SFA short and long trajectory interfering (lower). For the linear molecules we consider here, spectra are symmetrical by reflection with the molecular axis ($\theta=0$) and we therefore display only half of those spectra. Left panels correspond to the ${\rm H}_{2}$ molecule (with $3\times10^{13}~{\rm W}\cdot{\rm cm}^{-2}$ and $2850~{\rm nm}$ laser intensity and wavelength respectively) and right panels to the ${\rm H}_{2}^{+}$ molecular ion (laser parameters are the same as in Fig.~\ref{fig:HHG_reference_system}) model of potential~(\ref{eq:Potential}). For both systems, we focus on the part of the spectrum between $2.4Up+I_{p}$ and $3.17Up+I_{p}$ energy where only one short and one long trajectory (no multiple recollision) contribute to the spectrum in the SFA. For indication on each colormap we also display, with a continuous curve, the HHG energy associated with the singularity in the scattering cross-section (see Fig.~\ref{fig:Scattering_cross_section}).}
\end{figure*}

In figure~\ref{fig:Angle_resolved_spectrum}, we compare the accuracy of the two factorizations (middle and lower panels) in approximating the full quantum spectra (upper panels). More specifically, we display the intensity of odd harmonics (which gives the global envelope of the harmonic comb) between $2.4Up+I_{p}$ and $3.17Up+I_{p}$ energy, as the polarization angle $\theta$ is varied for the ${\rm H}_{2}$ molecule (left panels) and ${\rm H}_{2}^{+}$ molecular ion (right) models of potential~(\ref{eq:Potential}). Qualitatively, we see that both factorizations reproduce full quantum results very well. In particular, we see that all three panels present very similar oscillations patterns in photon energy and, as the polarization direction is varied, they all exhibit a local minimum that follows the singularity in the scattering cross-section (see black curves in the panels). For both factorizations though, we see that this local minimum is sharper than in the full quantum computation counterparts. This can be attributed to the fact that in full quantum simulations, the scattering cross-section is expressed in the temporal domain as in Eq.~(\ref{eq:Complex_dipole:factorization}) while it is directly expressed in the frequency domain for the factorization. In the former case, the singularity can be blurred by higher order effects when computing the Fourier transform to compute the HHG spectrum~(\ref{eq:HHG_spectrum}). On the quantitative level, in Fig.~\ref{fig:Angle_resolved_spectrum}, we observe better results using the reference system (middle panels) than with the direct factorization with SFA (lower). As discussed in section~\ref{sec:HHG_factorization}, we attribute this to the fact that the reference includes long-range Coulomb corrections to the electron dynamics away from the core whereas when choosing the SFA in the direct factorization we ignore such effects altogether.

Beyond the results of the factorization derived in this Article, the SCWF allows the identification of possible perspectives for improving the results of approximate predictions compared to full quantum simulations. For the reference system of Eq.~(\ref{eq:HHG_spectrum:Reference_atom:Factorization}) there is not much obvious room for improvement apart from potentially fine tuning the energy correspondence $E\approx E^{\prime}$ in the scattering cross-section. On the other hand, the direct factorization, with a direct access to the propagation step and underlying trajectories, leaves more perspectives for improvement. One such logical possibility is accounting for the energy Stark-shift~\cite{Etch10} in both the bound part of the wavefunction and instantaneous frequency. This formalism also leaves room for including laser induced bound state deformation~\cite{Spie13,Zhan14}.

\section*{Acknowledgements}

This work was supported by U.S. Department of Energy, Office of Science, Office of Basic Energy Sciences, under Award No.~DE-SC0012462.




\begin{thebibliography}{99}
\bibitem[{not()}]{note_Ip}
\bibinfo{note}{Because of the reduced one dimensional configuration we consider
  here, the ionization potential slightly varies with the polarization angle
  $\theta$. In numerical analyses, we account for this slight variation by
  shifting spectra in energy (frequency) accordingly. From the definition of
  the instantaneous frequency~(\ref{eq:Instantaneous_frequency}) we take this
  shift equal to the difference in the ionization potentials
  $Ip\left(\theta\right)-Ip^{\prime}$.}
\end{thebibliography}

\end{document}